\documentclass{aastex701}

\journalinfo{Submitted to ApJ}

\begin{document}

\title{Nuclear Isomers and Their Impact on Gamma-Ray Emission in Binary Neutron Star Mergers}

\author[orcid=0000-0003-1878-2445,gname=Maria,sname='Babiuc Hamilton']{M. C. Babiuc Hamilton}
\affiliation{Marshall University, Huntington, WV, 25755, United States}
\email[show]{babiuc@marshall.edu}  

\author[gname=Axel,sname=Gross]{A. P. Gross}
\affiliation{Theoretical Division, Los Alamos National Laboratory, Los Alamos, NM, 87545, United States}
\email{agross@lanl.gov}

\author[gname=Owen, sname='Odney']{O. T. Odney} 
\affiliation{Purdue University, West Lafayette, IN, 47907, United States}
\email{oodney@purdue.edu}

\begin{abstract}

The multi-messenger observations of GW170817 provided strong evidence that binary neutron star mergers are a site of heavy-element production through the rapid neutron capture process.
The decay of unstable $r$-process nuclei produces a significant number of $\gamma$-rays, which not only power the associated kilonova, but may also be observable in current and future $\gamma$-ray observatories, providing a direct probe of the nuclei synthesized. 
Current models of the emitted $\gamma$-rays link a nuclear reaction network with individual decay spectra. Typically, only the population of the ground state of the nuclei is tracked in the reaction network, and rapid de-excitation of the daughter nuclei is assumed when calculating the decay spectra. This may not be accurate in nuclei where there are longer-lived, high-energy nuclear isomers. In this work, we relax these assumptions, modeling the dynamic behavior of nuclear isomers for a select list of nuclei. These isomers are included as independent states in the network, with temperature-dependent effective transition rates between ground and isomeric states and $\beta$-feeding probabilities calculated using the nuclear level structure. We estimate the $\gamma$-ray flux from isomeric transitions, finding that the 743.3 keV line of Nb-97m and the 555.6 keV line of Y-91m may be detectable in $\gamma$-ray observatories such as COSI, AMEGO, and LOX at galactic distances. These results, calculated for a small fraction of nuclei with isomers, highlight the need for the robust inclusion of nuclear isomers in nuclear reaction networks, as well as careful modeling of the resultant $\gamma$-ray spectroscopy to characterize observational signals.

\end{abstract}


\section{\label{sec:intro}Introduction} 
The rapid neutron capture process ($r$-process) is a significant source of elements heavier than iron. Binary neutron star (BNS) mergers have long been proposed as sites of the $r$-process, and the landmark multi-messenger observations of GW170817 and its associated kilonova AT2017gfo \citep{arxiv:1710.05834, arxiv:1710.05446, arxiv:1711.03982} provided observational confirmation of these predictions. In particular, the observed optical and near-infrared peak has been interpreted as evidence of substantial lanthanide production.
There has been additional indirect identification of individual r-process elements in the merger ejecta. Strontium (Sr), a first peak r-process element, was identified in GW170817 \citep{arxiv:1910.10510}, while a spectral line associated with tellurium (Te), a second peak r-process element, was observed following GRB 230307A \citep{arxiv:2307.02098, arxiv:2308.00633}.

While the importance of these observations cannot be overstated, it is challenging to connect signatures of the light-curve to specific $r$-process physics. Optical and near-infrared data can be used to constrain bulk ejecta properties such as mass, velocity, and lanthanide fraction \citep{arxiv.1710.05858}, but they provide limited information about individual nuclei because the relevant observables are integrated over many decay chains.

Gamma-ray line spectroscopy offers a complementary probe to kilonova observations. The resolution of individual decay transitions provide a direct fingerprint of the nuclei present in the ejecta. 
With upcoming MeV instruments such as Compton Spectrometer and Imager (COSI) \citep{arxiv:1908.04334, arxiv:2109.10403, arxiv:2308.12362} and proposed missions such as All-sky Medium Energy Gamma-ray Observatory (AMEGO) \citep{arxiv:2101.03105} and Lunar Occultation eXplorer (LOX) \citep{arxiv:1907.07005}, this becomes a near-term observational opportunity. 
The correct interpretation of these observations require accurate prediction of $\gamma$-ray emission. Existing predictions (see e.g. \cite{arxiv:1511.05580, arxiv:1905.05089, arxiv:2107.02982, arxiv:2112.00772, arxiv:2204.13269, arxiv:2402.08214, arxiv:2402.11304, arxiv:2407.14762, arxiv:2409.15653, arxiv:2512.23354, arxiv:2510.08560}) compute the $\gamma$-ray spectrum by convolving r-process abundances with tabulated $\beta$- and $\alpha$-decay $\gamma$-lines from nuclear databases. 
This approach commonly relies on two implicit assumptions: 
(i) the r-process network follows ground-state species only;
(ii) decay tables used for the $\gamma$-ray spectrum fold isomeric transitions into the parent spectrum, with excited states populated from $\beta$-decay cascade into the ground state,
all emitted $\gamma$-rays being treated as instantaneous relative to the decay of the parent nucleus.

We first consider the impact of the ground-state approximation. This assumption fails when an isomer has a half-life comparable to or longer than the associated ground state. 
More than $700$ nuclei are known to host isomeric states with terrestrial half-lives exceeding 100 $\mu$s \citep{arxiv:2103.13133, arxiv:2208.01028}, and many along the r-process path have half-lives of seconds to years. The dynamical treatment of these isomers may lead to alterations of the decay timescale, or in rarer cases, the decay channels, which may modify the ejecta heating and thus the kilonova light curve \citep{arxiv:2103.09392,arxiv:1803.08335, arxiv:2001.10668}. 
The second assumption breaks down in multiple scenarios. First, isomers can directly decay via other channels, including $\beta$, $\alpha$, and fission, thereby bypassing the transition into the ground state. These decays may have significantly different emission properties than the decay of the corresponding ground state. In addition, the thermal history of the ejecta can modify the effective transition rates. At high temperatures, population of the isomer is thermally favorable, while at low temperatures, the ground state is favored. However, the effective transition rates are dependent on the allowed transitions and level structure. In some nuclei, intermediate states remain thermally accessible and mediate population transfer from isomer to ground, but in others, the relevant transition pathways become inaccessible as the ejecta cools, causing the isomer to decouple entirely from the ground state and evolve as an independent species. The standard tabulation of $\beta$-decay $\gamma$-lines assumes that isomeric states decay promptly to the ground state. However, if the isomeric state is long-lived, the same $\gamma$-line is emitted on a timescale set by the isomer half-life, not the parent $\beta$-decay half-life, and the standard pipeline therefore assigns an incorrect time dependence to the flux.

The significant impact of nuclear isomers on $r$-process nucleosynthesis has already been demonstrated. For example, Misch et al. \citep{arxiv:2011.11889}, showed that many nuclear isomers are ``astromers'', which evolve out of thermal equilibrium with the associated ground state and therefore must be treated dynamically as their own species in their network. Astromers become relevant within the first few seconds, well before $\gamma$-ray observations become feasible, but remain relevant for very long timescales, as they are often populated directly from $\beta$-decay and are therefore populated out of thermal equilibrium. In earlier work \citep{arxiv:2402.06498}, we advanced that the r-process produces nuclei in metastable, isomeric states, forming a potentially distinct source of $\gamma$-rays. In this work, we extend this by modeling the contributions to $\gamma$-rays from nuclear isomers in detail. 
While isomers that directly $\beta$-decay also produce a $\gamma$-ray spectrum distinct from that of the corresponding ground state, we focus here on isomers that decay through isomeric transition, generating narrow $\gamma$-ray lines that provide promising observational signatures. 
We restrict our analysis to a small subset of 12 isomers, computing the effective transition rates between ground and isomeric states using two established approaches, validating the results against the well-studied Al–26m isomer data \citep{arxiv:2010.15238, meyer}. 
We then construct nuclear input datasets and extend the PRISM reaction network \citep{arxiv:1508.07352} to explicitly track the time-dependent evolution of these isomeric species. We apply this framework to 30 ejecta trajectories from numerical relativity simulations of BNS mergers, representative of the range of nucleosynthesis which occurs in the disk \citep{arxiv:2311.05796}, finding that six of the isomers reach non-trivial abundances. For these nuclei, we analyze the potential for $\gamma$-ray observation, finding that the 555.6 keV decay line of Y-91m and 743.3 keV decay line of Nb-97m are dominant contributors to the $\gamma$-ray flux at their respective energies and may be detectable by $\gamma$-ray observatories such as COSI, AMEGO, and LOX at galactic distances. With these findings, we argue that the standard modeling pipeline is insufficient to accurately characterize the $\gamma$-ray emission from $r$-process nucleosynthesis, and that careful treatment of isomers in reaction networks and the resulting spectroscopy is required.

The remainder of the paper is organized as follows: 
In \S\ref{sec:selection}, we describe the selection of isomer candidates from the chart of nuclides, combining a priori criteria ($\gamma$-ray energy, decay timescale, and r-process accessibility) with posterior validation through the reaction network to ensure the isomers are sufficiently populated to have potential observational consequences.
In \S\ref{sec:theory}, we develop the theoretical framework for computing temperature-dependent effective transition rates between isomeric and ground states, including the role of $\beta$-feeding and the use of Weisskopf estimates when experimental data are incomplete. 
\S\ref{sec:implementation} describes the extension of the PRISM reaction network to evolve isomers as independent species and its application to post-merger ejecta trajectories from a three-dimensional magnetohydrodynamics general-relativistic simulation of a BNS remnant accretion disk.
In \S\ref{sec:observability} we calculate the $\gamma$-ray flux associated with isomeric transitions and compare to other relevant sources of $\gamma$-rays, including the thermal background and $\beta$-decays of other unstable $r$-process nuclei. We model the escape of these $\gamma$-rays via Compton opacity under two escape geometries that bracket the physical regime and compare the estimated $\gamma$-ray flux to the sensitivities of $\gamma$-ray observatories. 
In \S\ref{sec:discussion} we discuss our findings, highlighting the importance of the dynamical treatment of isomers for $\gamma$-ray spectroscopy. Finally, \S\ref{sec:conclusion} summarizes our main conclusions.

Our implementation of temperature-dependent isomeric transition rates data calculation and its conversion to PRISM input files is publicly available online at \url{https://doi.org/10.5281/zenodo.20433267}.
The processed nuclear data files and gamma-ray spectra analysis notebooks are available upon request from the authors.

\section{\label{sec:selection}Selection of Observable Isomer Candidates}
We identify isomers that may produce observable $\gamma$-ray signals by imposing a sequence of physically motivated filters. 
First, we apply a priori cuts based on tabulated nuclear properties. We then run nucleosynthesis calculations with the a priori candidates across 30 post-merger nucleosynthetic trajectories and apply posterior cuts based on the output.  

\subsection{A-priori Selection}
An observable isomer candidate must be produced by the r-process, persist on observable timescales, and decay through a distinct $\gamma$-ray line with energy above the thermal background. 
Although isomers may also directly decay through other channels, emitting spectra distinct from the corresponding ground states, we focus on isomeric transitions to the ground state because they produce identifiable $\gamma$-ray lines that can be directly linked to specific nuclei.
We therefore require that the isomer decays via isomeric transition (IT) with branching ratio greater than 99\%, based on terrestrial evaluations \citep{arxiv:2208.01028}.  These isomers must also be produced in $r$-process scenarios. This requirement restricts the neutron number to N $\ge$ 50, demands the existence of at least one $\beta$-decay parent, and excludes proton-rich nuclei.

We then impose a lower bound of 0.3 MeV on the energy level of the isomer. This places the emitted photon within the 0.2-5 MeV sensitivity range of current and planned instruments (COSI, AMEGO, LOX). We also require a minimum half-life of 1 ms. This excludes states that decay within the prompt-cascade regime assumed in standard pipelines and cannot maintain significant isomeric abundances in reaction networks. These two filters are physically related, because low-energy IT transitions tend to be fast.

We next require the half-life of the $\beta$-decay parent to fall between minutes and approximately one year. 
Parents with sub-minute half-lives decay well before the observation window opens, while parents with multi-year half-lives do not decay during the main observational window.
In addition, any isomer populated only at such late times would receive little population during the epoch of interest, minimizing its contribution to the observable signal.

Applying these cuts reduces the initial set of $50$ candidates to the $12$ isomers which we have implemented in our reaction network modeling.

\subsection{Posterior Selection}
We follow up our selection by examining which of these candidates are realized in $r$-process trajectories. First, we require that the isomer abundance exceeds a minimum numerical threshold ($Y_{\rm iso}^{\rm max} \ge 10^{-12}$) in at least one of the 30 trajectories. This criterion removes six candidates that are minimally populated by the r-process. 
We then evaluate the isomer to ground abundance ratio $Y_{\rm iso}/Y_{\rm g}$. 
Candidates with $Y_{\rm iso}/Y_{\rm g} \ge 1$ are classified as {\em most promising}, since the isomer population exceeds that of the ground state. We retain candidates with $Y_{\rm iso}/Y_{\rm g} < 1$, as their contribution to the spectrum may remain non-negligible if the line produced is particularly prominent. 
These nuclei are located near the neutron magic numbers (N $\sim$ 50 , 82 , 126), in the so-called \emph{islands of isomerism} \citep{arxiv:2208.01028}. 
Although these nuclei do not overlap with the major $r$-process abundance peaks, they are produced in non-negligible amounts and may contribute to the $\gamma$-ray spectrum. 
This selection explores only a limited set of candidate isomers and therefore underestimates all possible isomeric contributions.

In the following analysis, we examine the contributions from the six isomers we have identified, which is sufficient to demonstrate that dynamically populated isomers can produce observable $\gamma$-ray lines.

The 12 a-priori selected isomers are listed in Table~\ref{tab:isomers}, together with their excitation and de-excitation energies, half-lives, $\beta^-$ decaying parent (with parent half-life), maximum per-trajectory abundance and abundance ratio. The posterior selection yields six candidate isomers: 
Y-91m, Nb-97m, Zr-90m, Yb-175m, Lu-179m, and Tl-207m.

\begin{deluxetable*}{rllllll}[ht!]
\tablecaption{Selected r-process isomers for simulations}
\label{tab:isomers}
\tablehead{
\colhead{Isomer} & 
\colhead{E$_\gamma$(keV)} &
\colhead{T$_{1/2}$} & 
\colhead{Parent (T$_{1/2}$)} &
\colhead{$Y_{\rm iso}^{\rm max}$} & 
\colhead{$\left ( Y_{\rm iso}/Y_{\rm g} \right )^{\rm max} $}
}
\startdata
Y-91m   & 555.6  & 49.7 min & Sr-91 (9.65 h)    & $9.863\times 10^{-6}$ & $2.443 $\\
Zr-90m  & 2319.0 & 809 ms   & Y-90 (64.05 h)    & $1.001\times 10^{-6}$ & $1.123\times 10^{-1}$\\
Nb-97m  & 743.4  & 58.7 s   & Zr-97 (16.7 h)    & $9.319 \times 10^{-8}$ & $2.198$\\
Xe-132m & 538.2  & 8.39 ms  & I-132 (2.295 h) & $< 10^{-12}$ & -- \\
Xe-135m & 526.5  & 15.3 min & I-135 (6.58 h)    & $< 10^{-12}$ & -- \\
Cs-135m & 846.1  & 53 min   & Xe-135 (9.14 h)   & $< 10^{-12}$ & -- \\
Yb-175m & 515.0  & 68.2 ms  & Tm-175 (15.2 min) & $9.303 \times 10^{-10}$ & $1.066\times 10^{-1}$\\
Lu-179m & 592.4  & 3.1 ms   & Yb-179 (8.0 min)  & $4.523\times 10^{-11}$ & $1.964\times 10^{-3}$\\
Hf-178m & 587    & 31 yrs   & Lu-178 (28.4 min) & $< 10^{-12}$ & -- \\
Tl-206m & 1021.5 & 3.74 min & Hg-206 (8.32 min) & $< 10^{-12}$ & -- \\
Tl-207m & 997.1  & 1.33 s   & Hg-207 (2.9 min)  & $2.703\times 10^{-9}$ & $2.459\times 10^{-2}$\\
Pb-207m & 1063.6 & 806 ms   & Tl-207 (4.77 min) & $< 10^{-12}$ & -- \\
\enddata
\end{deluxetable*}

\section{\label{sec:theory}Theoretical Framework for Isomeric Transition Rates}
We present below a robust framework for calculating the effective transition rates, which take into account the entire nuclear level structure to obtain the thermally mediated rates between ground and isomeric states. To this end, we adopt the method introduced in \citep{meyer} and extended in \citep{arxiv:2010.15238}.

We consider a nucleus with $n$ discrete energy levels, assign to each level an index $i$, such that the ground state corresponds to $i=0$ (denoted $g$) and the isomeric state to $i=m>g$. 
The transition rates between levels $i \rightarrow j$ is denoted by the index pair $ij$, replaced by $hl$ for a generic downward transition and $lh$ for a generic upward transition.
When the long-lived states are treated without specifying whether they are ground or isomeric, we refer to them as end states and denote their indices by $E$, or $A$ (initial end state) and $B$ (final end state).

\subsection{Constructing gamma-transition Effective Rates}

We start from the downward transition rate $\lambda_{hl}$ (either spontaneous or stimulated), which is calculated using the half-life of the higher energy level $T_{1/2}(E_h)$ weighted by its transition probability $w_{hl}$:
\begin{equation}
    \lambda_{hl} = w_{hl}\frac{\ln 2}{T_{1/2}(E_h)}.
    \label{eq:down}
\end{equation}
The weighted transition probability is estimated from the measured $\gamma$-ray intensity $I_{\gamma, h}$ along the decay path:
\begin{equation}
w_{hl} = \frac{I_{\gamma,h}}{\sum_h I_{\gamma,h}}.
\label{eq:weight}
\end{equation}

We then compute the upward induced transition rate $\lambda_{lh}$, assuming a thermal photon bath described by a Planck distribution at temperature $T$:
\begin{equation}
    \lambda_{lh} = \lambda_{hl} \frac{g_h}{g_l} e^{-\Delta E_{hl} / k_B T}.
\end{equation}
Here $g_i = 2 J_i +1$ are the degeneracies associated with the spins $J_i$, $\Delta E_{hl} = E_h - E_l$ represent the energy differences between levels, and $k_B$ is the Boltzmann constant.

Next, we define the step probability of transitioning from a state $i$ to a state $j$, or the branching ratio as: 
\begin{equation}
   b_{ij} = \frac{\lambda_{ij}}{\lambda_{i}},
\end{equation}
where $\lambda_{i}= \sum_{k\ne i} \lambda_{i k}$ represents the total rate for depopulating level $i$. 
Since a state cannot transition to itself, $b_{ii}=0$.
These probabilities form a hollow matrix $\mathbf{b}_{IJ}$, with each row normalized such in that row $\sum_j b_{ij} = 1$.

From this point, the two methods diverge slightly. In \citep{meyer} (hereafter referred to as {\bf M1}), the inversion of the branching matrix $\mathbf{b}_{IJ}$ is avoided by constructing a partial-sum matrix $\mathbf {F}$ which has the same elements as $\mathbf{b}_{IJ}$ but reduced order $(n-2)\times(n-2)$, because the ground and metastable (isomeric) states are excluded:
\begin{equation}
    F_N = I + F + F^2 + ... + F^{N-1}.
\end{equation}
Here $N = n-1$ denotes the number of excited states.
The branching ratio vectors corresponding to transitions into and out of a long-lived or end state (labeled $E\in{g, m}$) are then defined by excluding both the ground and isomeric levels from the summation, such that $j\ne {g, m}$:
\begin{equation}
f_E^{\rm in} = f_{j E} = b_{j E}, 
\quad \texttt{and} \quad 
f_E^{\rm out} =  f_{E j} = b_{E j}.
\end{equation}
Lastly, the {\bf M1} method calculates the effective transition rate from an initial to a final long-lived state with:
\begin{equation}
\Lambda_{AB} = \lambda_{A} (f_A^{\rm out})^T F_N f_B^{\rm in}.
\label{eq:LambdaM1}
\end{equation}

The second approach, {\bf M2}, presented in \citep{arxiv:2010.15238}, introduces the quantity $P_{iB}$, defined as the probability for the system to reach the final end state $B$ by a random walk from an intermediate level $i$ before returning to an initial state $A$.
In this formulation, $P_{iB}$ is obtained by summing over all accessible states to determine the probability of reaching a particular state through intermediate steps:
\begin{equation}
P_{i B} = b_{iB}+\sum_{j\ne A, B}b_{ij}P_{j B}.
\end{equation}
The corresponding system of linear equations can be expressed in matrix form and solved by inversion, the resulting matrix being shown in \citep{arxiv:2010.15238} to be invertible:
\begin{equation}
\mathbf {P}_{IB} = (\mathbf{I} - \mathbf{b}_{IJ})^{-1}\mathbf{b}_{IB}.
\label{eq:vectorP}
\end{equation}
The effective transition rate from an initial end state $A$ to a final end state $B$ is calculated in {\bf M2} as:
\begin{equation}
\Lambda_{AB} = \lambda_{AB} + \sum_{i\ne{A,B}}  \lambda_{Ai} P_{iB}.
\label{eq:LambdaM2}
\end{equation}

The principal difference between the two methods lies in their treatment of intermediate transitions: {\bf M2} solves the system of equations directly, providing an exact solution, while {\bf M1} approximates multistep transitions by truncating the matrix $F_N$, where $N$ is chosen high enough to ensure convergence.

\subsection{Adding Beta-decay and Feeding Effective rates}
The effective transition rate $\Lambda_{AB}$ provides direct insight into how the effective lifetime of an isomeric state varies with temperature in a thermal environment due to thermally mediated $\gamma$ transitions. However, an isomer may also decay or be fed through intrinsic channels, such as $\beta$-decay, which are generally considered temperature-independent.
To account for this additional decay channel, we calculate the effective $\beta$-decay rate of the nucleus in an end state (ground or isomer) following the time-dependent procedure described in \citep{meyer}.

Assuming a Boltzmann distribution at temperature $T$ between the abundance of a nuclear species in an intermediate level $i$ and the end level $E \in\{A, B\}$, we define the reverse ratio as:
\begin{equation}
R_{Ei} = \frac{Y_i}{Y_E} = \frac{\lambda_{Ei}}{\lambda_{iE}} = \frac{2 J_i + 1}{2 J_E +1}e^{(E_E - E_i)/k_B T}.
\end{equation}
The $\beta$-decay process is characterized by transition weights that specify whether the decay affects the ground or the isomeric state:
\begin{equation}
w_{iE} = P_{iE}R_{Ei}~\textrm{if}~ i\neq\{A,B\},~~\textrm{and}~ w_{iE} = \delta_{iE}~\textrm{if}~i = \{A, B\},
\end{equation}
such that the abundance in the intermediary level $i$ is:
\begin{equation}
Y_{i} = w_{iA} Y_A + w_{iB} Y_B,,
\end{equation}
These weights are summed as $W_E = \sum_i w_{iE}$, and the total abundance is written as the sum of weighted abundances of two end states:
\begin{equation}
Y_{T} = W_A Y_A + W_B Y_B,
\end{equation}
The total effective $\beta$-decay is then expressed as a linear combination of $\beta$-decay rates from intermediary excited states $i$, each weighted by its fractional population:
\begin{equation}
\Lambda_{T \beta} = \frac{\sum_i {\lambda_{i \beta} Y_i}}{Y_T} 
= \frac{\sum_i {\lambda_{i \beta} w_{iA}}}{W_A} \frac{Y_A}{Y_T} + \frac{\sum_i {\lambda_{i \beta} w_{iB}}}{W_B} \frac{Y_B}{Y_T},
\end{equation}
where $\lambda_{i \beta}$ denotes the $\beta$-decay rate out of level $i$.
The individual $\beta$-decay rates of the end-state channels are:
\begin{equation}
\Lambda_{E \beta} = \frac{\sum_i {w_{iE} \lambda_{i \beta}}}{W_E}.
\label{eq:beta}
\end{equation}

In the r-process environment, isomeric states are also populated through $\beta$-feeding from parent nuclei that decay into excited daughter levels, which cascade to the isomeric or ground states.
To calculate the effective rates by which the isomeric and ground states are populated through $\beta$-decay, we define the probability that an end state $E$ is populated:
\begin{equation}
P_E = \sum_i {w_{i,\beta} P_{i E}},
\end{equation}
where $P_{i E}$ is the cascade fraction from level $i$ to end-state $E$, determined by the probability vector $\mathbf {P}_{IE}$ computed with the same matrix formalism (eq.\ref{eq:vectorP}), and $w_{i\beta}$ the normalized $\beta$-decay intensities populating daughter level $i$:
\begin{equation}
w_{i,\beta} = \frac{I_{\beta,i}}{\sum_i I_{\beta,i}}.
\end{equation}
 The two end-state probabilities satisfy $P_g + P_m = 1$. The effective transition rate $\Lambda_{E \beta}^{\rm feed}$ is then given by:
\begin{equation}
\Lambda_{E}^{\rm feed} = \lambda_{\beta}^{\rm parent} P_E,
\label{eq:feed}
\end{equation}
where $\lambda_{\beta}^{\rm parent} = \ln (2) / T_{1/2}^{\rm parent}$ is the $\beta$-decay rate of the parent into the daughter isotope.
Physically, this reflects how higher temperatures populate intermediate excited levels in the daughter nucleus, which then cascade preferentially to the metastable state rather than directly to the ground state. 
This effect partially compensates for the rapid isomer to ground destruction rates. Even as the isomer is thermally driven toward the ground state, $\beta$-feeding from longer-lived parents continues to refill the isomeric state preferentially at high temperatures.

\subsection{Supplementing with Theoretical Weisskopf Estimates}
These calculations require the half-lives $T_{1/2}$, relative intensities $I_{\gamma}$ and spins $J$ for each decay path, the $\beta$-decay rates of ground and isomeric levels, and the $\beta$-feeding probabilities into each daughter state.
When available, we use experimental data, taken from \emph{ENSDF} and \emph{IAEA} databases \citep{ENSDF, IAEA}. However, the experimental nuclear data are often incomplete, so we supplement it with theoretical Weisskopf estimates, which provide order-of-magnitude estimation of the transition rates \citep{arxiv:1509.09101, arxiv:1803.08335}.
The details of how we incorporate the theoretical and experimental data are outlined below.

A transition is allowed if it satisfies both angular momentum and parity conservation laws, determined by the difference in spin $J$ and parity $\pi$ (even $+$ or odd $-$) between the initial and final states $(J_i, \pi_i) \rightarrow (J_f, \pi_f)$ of a nucleus.
The angular momentum $L$ carried by the emitted $\gamma$-photon must satisfy the conservation law:
\begin{equation}
|J_i - J_f | \le L \le  |J_i + J_f |,~~L\ge 1.
\label{eq:Lrule}
\end{equation}
The transition is electric if $\pi_i \pi_f = (-1)^L$ and magnetic if $\pi_i \pi_f = (-1)^{L+1}$. The value of $L$ determines the multipolarity of the emitted $\gamma$-ray, with the lowest $L$ satisfying both rules defining the allowed transition.

The Weisskopf estimates for the decay rates are expressed as in \citep{Blatt}:
\begin{equation}
\lambda^W_{hl} =\frac{8 \pi (L+1)}{\hbar L \left ((2L+1)!!\right )^2}\left ( \frac{E_{\gamma}}{\hbar c}\right )^{2L+1} B(E, M),
\end{equation}
where $E_{\gamma} =\Delta E_{hl}$ is the energy released in the decay.
The corrections $B(E, M)$, known as multipole matrix elements, depend on the multipolarity of the transition.
Since direct evaluation of these terms can be cumbersome, we use tabulated Weisskopf estimates for typical electromagnetic transitions up to $L = 5$:  
\begin{equation}
\lambda^{E L} = C_{E}( L) A^{2 L/3} Q^{2 L+1}
\quad \rm and \quad
\lambda^{M L} = C_{M}( L) A^{(2 L-2)/3} Q^{2 L+1}.
\label{eq:Weisskopf}
 \end{equation}
where $Q$ is $E_{\gamma}$ in MeV, and the constants $C_{E,M}(L)$ are given in \citep{vanDommelen}.
We exclude levels with unknown spin-parity from our nuclear level structure to avoid additional uncertainties.

\subsection{\label{sec:rate}Implementation of the Rate Formalism}
We begin by collecting four experimental nuclear datasets \citep{ENSDF, IAEA} 
(i) energy levels, including spin, parity, and half-life, (ii) $\gamma$-ray transition data, including transition energies and intensities, (iii) $\beta$-decay rates and (iv) $\beta$-decay rates of the parent, including intensities of the decay paths and daughter level energies.

The raw data require substantial pre-processing, including removal of extraneous headers and metadata, normalization of numerical formats, and standardization of inconsistent representations. 
We therefore developed a dedicated cleaning script that produces both a minimal version and a fully cleaned dataset for transparency.

For levels lacking experimental half-life information, we estimate $\gamma$-decay rates using Weisskopf (eq.\ref{eq:Weisskopf}) to fill in missing data up to level 60.
The transition multipolarity is determined from angular-momentum selection rules (eq.\ref{eq:Lrule}) and the relative parity of the initial and final states. For each level with missing data, the script finds all allowed transitions to lower-lying states, computes the corresponding partial decay rates, sums them, and derives an estimated half-life.
This procedure yields a complete and self-consistent set of downward transition rates, which are then used to compute the total transition rate via (eq.\ref{eq:down}), with the weighted transition probabilities given by (eq.\ref{eq:weight}).

We next implement the temperature-dependent effective transition rates using the {\bf M2} formalism described in \S\ref{sec:theory} and summarized in (eq.\ref{eq:LambdaM2}).
{\bf M2} is exact and serves as our default for production runs.
However, it requires inverting the level-coupling matrix and becomes numerically unreliable when the matrix is ill-conditioned, typically at low temperatures, where the populations of intermediate states span many orders of magnitude. 
We therefore implement in parallel the earlier {\bf M1} power-series formalism, which is robust in this regime but truncates the level coupling at finite order. 
We use {\bf M1} as a fallback whenever the {\bf M2} matrix condition number exceeds $10^{12}$. For well-conditioned systems, the two methods agree to machine precision, providing a cross-validation of our numerical implementation.

We then implement temperature-dependent effective $\beta$-decay rates from each nuclear state (ground or isomer) following (eq.\ref{eq:beta}), under the assumption of thermal equilibration between ground-state and isomeric populations.

Finally, we evaluate the temperature-dependent $\beta$-decay feeding probabilities from a parent nucleus to the ground or isomeric state of the daughter nucleus, as defined in (eq.\ref{eq:feed}).
This enforces probability conservation and captures the temperature dependence of ground-to-isomer branching ratios as cascade pathways become thermally accessible.

In Appendix \ref{sec:validation} we present tests of the calculations described above, while Appendix \ref{sec:application} illustrates their application to the selected set of representative isomers produced within the ejecta trajectories.

\section{\label{sec:implementation}Network Implementation and Ejecta Model}

\subsection{PRISM Network with Isomer Tracking}

Our nucleosynthesis calculations are performed with version 1.6.0 of the Portable Routines for Integrated nucleoSynthesis Modeling (PRISM) code \citep{arXiv:2008.06075}, which includes charged-particle reactions, neutron capture, photodissociation, $\beta$-decay, delayed neutron emission, and fission \citep{arXiv:1706.07504, arXiv:1802.04398, arXiv:1911.06344}.

PRISM tracks nuclear species as $(Z, N)$ pairs and therefore does not easily support isomeric states. To overcome this limitation, we extend the network to treat isomers as independent species with different $(Z,N)$ numbers, adopting a simple encoding scheme in which an isomer of nuclide $(Z,N)$ is represented as $(Z+1000, N)$. This approach allows the dynamic evaluation of ground and isomeric states using the effective transition rates determined above, requiring only minimal changes to the PRISM source code. Modifications are confined to a small set of routines (control, network size, nuclear data handling, output, and time progression). Mass conservation is verified using the built-in mass-loss diagnostic and full backward compatibility is maintained.

We provide as initial data the temperature-dependent effective transition rates between the ground and isomeric state, as well as effective $\beta$-decay rates to both the ground and isomeric state, which override the standard decay rate. 
For $\beta$-feeding rates we consider two treatments: temperature-independent, with the rates fixed at their optimal values,
and temperature-dependent. 
We compare the two approaches in Appendix \ref{sec:sensitivity} and adopt the temperature-dependent treatment as our fiducial model, since it captures the correct thermal redistribution of feeding across daughter levels.

Experimental $\beta$-feeding data are often incomplete. In particular, high-energy transitions that cascade through intermediate levels are frequently missing, so the tabulated feeding fractions do not sum to unity. We close this budget by assigning the missing fraction to the ground state. This reflects the expectation that unresolved cascades are more likely to terminate in the ground state than in a metastable configuration. 
The same convention is applied consistently in both the temperature-dependent implementation adopted here and in the  temperature-independent treatment.

\subsection{\label{sec:trajectory}Post-merger Ejecta Trajectories}

Numerical simulations show that the production of heavy r-process elements in BNS merger ejecta strongly depends on the angular distribution of the outflow. Polar regions eject early lanthanide-poor material dominated by lighter nuclei ($A \le 140$), while equatorial regions produce late-time lanthanide-rich ejecta containing heavier elements ($A > 140$) \citep{arxiv:1710.05463, arxiv:1907.04872, arxiv:2409.18185}. The thermodynamic trajectories used in this work are drawn from the general relativistic neutrino radiation magnetohydrodynamic simulation of a post-merger black hole--accretion disk system by \citet{arxiv:2311.05796}, specifically the \emph{b10} configuration (initial plasma $\beta = 10$).
The \emph{b10} model spans a broad range of electron fractions ($0.1 \le Y_e \le 0.45$), and ejects both lanthanide-poor ($A \le 140$) and lanthanide-rich ($A \ge 140$) material.
This model produces the largest ejecta mass among the three simulated magnetic field strengths and therefore provides the most favorable conditions for detectable $\gamma$-ray signals.

These simulations follow $\sim 1.5 \times 10^6$ Lagrangian tracer particles over $127$ ms of post-merger evolution. 
Among these, $195,288$ tracers satisfy the extraction criteria of reaching a radius of $250$ gravitational radii ($\sim 10^3$ km) with a positive Bernoulli parameter, corresponding to a total unbound ejecta mass of $7.31 \times 10^{-3} M_{\odot}$.
From these unbound tracers, only a representative subset of $30$ was selected in \citet{arxiv:2311.05796} for the purpose of subsequent nuclear sensitivity studies.
The ejecta were binned in radial velocity and electron fraction, restricting the sample to $v_r < 0.2~c$, and the most massive tracer from each bin was retained.
This reduced set accurately reproduces the integrated final r-process abundance pattern of the full \emph{b10} ejecta and carries a total scaled mass of $6.84 \times 10^{-3} M_{\odot}$, corresponding to $\sim 94\%$ of the total ejecta mass. The remaining $6\%$ originate from a small high-velocity ejecta component with $v_r > 0.2~c$ that is not sampled by the selection procedure. 
The masses assigned to individual tracers are obtained in post-processing and represent relative mass weights within the $(v_r, Y_e)$ ejecta distribution rather than the physical masses of individual fluid elements. 
Beyond the merger simulation the trajectories are extrapolated assuming homologous expansion up to $10^3$~s. For each tracer, we use the complete thermodynamic history
and begin the network calculation at time $t_{10}$, at which the temperature last drops below $10$~GK ($T_9 = 10$). 
We evolve the network to a final time of $5.184 \times 10^6$~s (thw months).

Because the 30 tracers were selected to sample the ($v_r$, $Y_e$) distribution rather than the angular structure of the ejecta, their angular positions do not reproduce the geometric mass distribution reported in \citep{arxiv:2311.05796}. 
We nonetheless use this set as a representation of the nucleosynthesis in BNS mergers and retain the angular classification of \citet{arxiv:2311.05796} with $\theta_{\rm ex}$ measured from the equatorial plane (polar: $\theta_{\rm ex} \ge 45^{\circ}$); intermediate: $15^{\circ} < \theta_{\rm ex} < 45^{\circ}$; equatorial: $\theta_{\rm ex} \le 15^{\circ}$).
The ejection angle correlates with the thermodynamic conditions that influence the r-process nucleosynthesis, and therefore the resulting isomeric populations. However, we highlight that this correlation between thermodynamic evolution and angle does not imply that these tracers reproduce the true ejecta mass distribution as a function of angle, due to the selection effects in the tracer sample. The accurate calculation of direction-dependent fluxes would require summation over all of the tracers from the simulation. For these reasons, we adopt a spherically symmetric ejecta model in the observability analysis presented in \S\ref{sec:observability}.

\begin{figure}[ht!]
\plotone{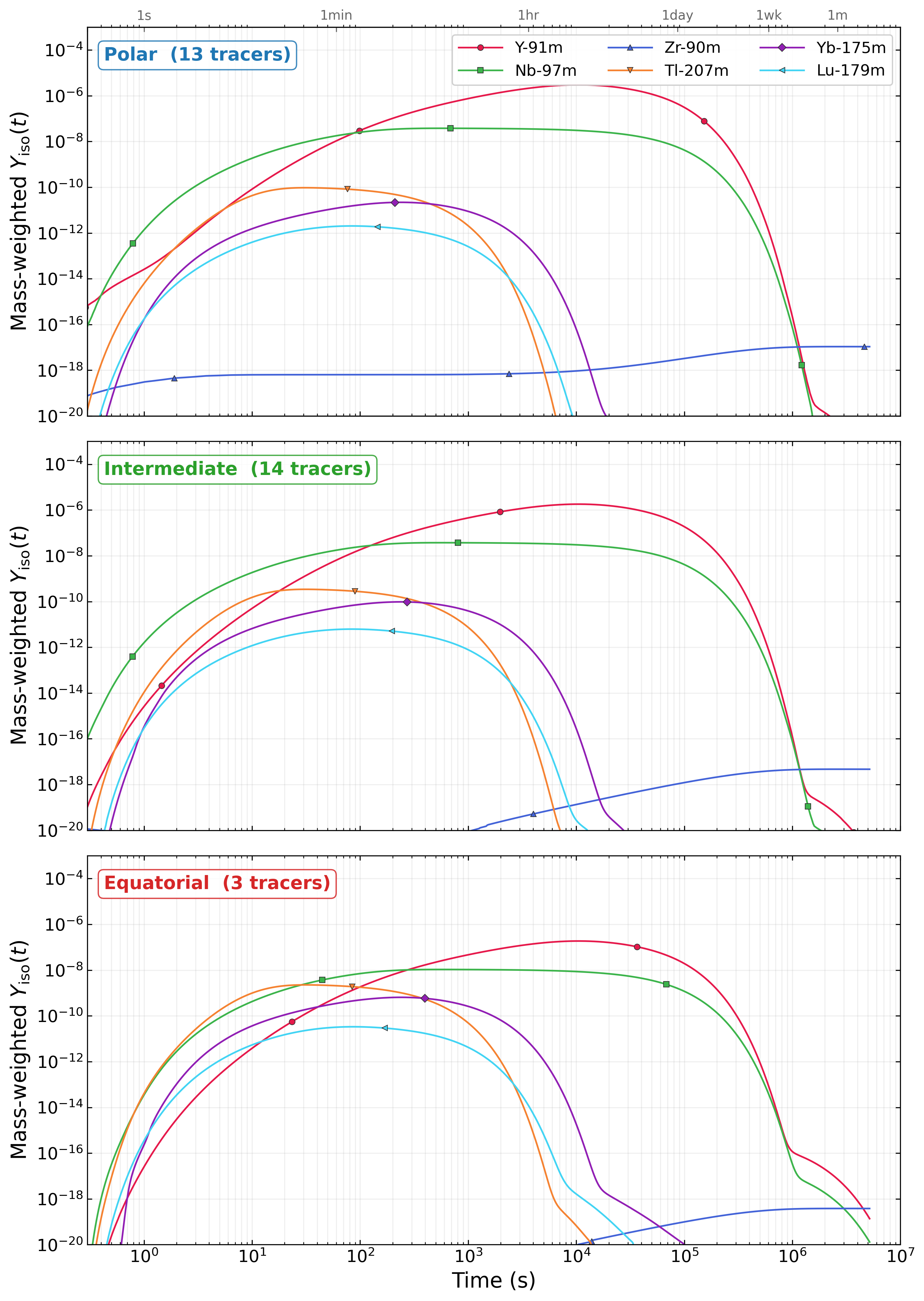}
\caption{Isomeric abundances for the isomers realized in the network above the numerical threshold $\ge 10^{-12}$. \label{fig:Figure1}}
\end{figure}

Figure \ref{fig:Figure1} shows the mass-weighted abundances of the six isomers  whose peak abundance exceeds our selection threshold ($Y\ge 10^{-12}$ in at least one of the 30 tracers). The angular differences are most pronounced for Y-91m abundances, which show an approximately two-order-of-magnitude decrease toward the equator, while the remaining isomers exhibit only modest variations.

\section{\label{sec:observability}Isomeric Signatures Observability}

Isomeric transitions produce strong $\gamma$-ray lines at well-defined energies, and may therefore provide clean spectral signatures which can be observed. To assess observability, we compute the $\gamma$-ray spectra from the selected isomers and compare them against two potential backgrounds: the thermal background of the ejecta and the significant amount of $\gamma$-rays which occur from the $\beta$-decay of other nuclei. To obtain the observed flux at a distant detector, we consider integration over the ejecta mass, the geometric dilution with distance $D$, and and the geometry-dependent escape fraction, which gives the probability that a photon of energy $E_{\gamma}$ escapes the expanding ejecta at time $t$.

\subsection{Geometry and photon escape fraction}
We estimate the asymptotic velocity of each trajectory by identifying the phase of homologous expansion, where $R(t) \approx v \cdot t$, and the density evolves as $\rho \propto t^{-3}$.
Thus, the quantity $\rho t^3$ is constant, and the ejecta velocity becomes:
\begin{equation}
v = \left (\frac{3 M_i}{4 \pi \rho(t) t^3} \right )^{1/3}.
\end{equation}

To identify the homologous phase for each tracer, we use a local flatness criterion, by computing $d \log (\rho t^3)/d \log t$ and selecting connected time intervals where this slope remains close to zero. These intervals correspond to candidate homologous plateaus.
Among all such plateaus, we select either the longest or the latest one, as it best represents the asymptotic coasting regime after transient effects have decayed, and calculate the median velocity over this interval. The resulting velocities span values from $0.003c$ to $0.145c$. 
The intermediate trajectories exhibit the widest velocity range, while the equatorial ejecta remains systematically slower.
This is consistent with the upper limit of $v_r < 0.2c$ imposed by the selection of representative tracers, and roughly in line with the literature \citep{arxiv:1905.05089}. High-velocity ejecta, not considered here, would contribute additional early-time emission with shorter escape timescales.

Because the 30 tracers are selected for nucleosynthetic fidelity rather than angular coverage, we do not decompose the ejecta into angular regions. Instead, we treat the outflow as a spherically symmetric shell and assign a single effective column mass to all trajectories of $M_{\rm total} = 6.84 \times 10^{-3} M_{\odot}$ corresponding to the total mass of the 30 tracers. 
This is conservative for detectability: in a structured ejecta, photons in the polar regime would traverse less material and escape faster.

The escape fraction depends on both the geometry and the column density traversed by photons. Since the ejecta morphology is not precisely known, we evaluate the escape under two idealized geometries and use them as a bracket on the observable signal.
We compute the escape fraction for a given optical depth $\tau(E,t)$ under the limiting geometries.

As first geometry, we choose the uniform-slab model, which has been used in previous kilonova modeling \citep{arxiv:1605.07218} and assigns the full column depth to every photon, therefore representing a pessimistic limit for a quasi-isotropic medium. This choice is appropriate when the density structure forces photons to traverse the maximum path length.
In this model, $\gamma$-rays escape along a single line-of-sight column of uniform density, and the escape fraction is defined as:
\begin{equation}
f_{\rm esc}^{\rm slab}(E, t) = e^{-\tau(E, t)}.
\end{equation}

The second geometry is the classic radiative-transfer escape-probability model for a homogeneous sphere \citep{Osterbrock}, in which photons are emitted and absorbed throughout a uniformly emitting sphere.
In this model, emission occurs throughout the ejecta volume rather than along a single line of sight, and photons emitted near the surface encounter smaller optical depths than those produced in the interior. The escape fraction is:
\begin{equation}
f_{\rm esc}^{\rm sphere}(E, t) = \frac{3}{8\tau^3} \cdot \left [2\tau^2 - 1 + (1 + 2\tau) e^{-2\tau} \right ].
\end{equation}

The two models behave differently in the optically thick regime. For ($\tau \gg 1$), the slab escape fraction falls exponentially as $e^{-\tau}$, while the sphere escape fraction scales as $\tau^{-1}$. 
As a result, the sphere model predicts higher escaping flux when the ejecta is still opaque. 
In the optically thin limit ($\tau \ll 1$), both models converge to $f_{\rm esc}\rightarrow 1$.

For the optical depth we use the expression:
\begin{equation}
\tau(E, t) = \kappa(E) \cdot \Sigma(t),
\end{equation}
where $\kappa(E)$ is the Compton opacity and $\Sigma(t) = 3 M_{\rm eff} / (4 \pi  R(t)^2)$ corresponds to the central line-of-sight column of a uniform-density sphere of mass $M_{\rm eff}$
and radius $R(t)$ along the escape path of the photon.

We use these two prescriptions as representative escape geometries that bracket the sensitivity of the predicted $\gamma$-ray emission to radiative transport assumptions.
The slab geometry generally produces more suppressed and delayed observable emission, while the spherical Osterbrock prescription yields earlier and brighter escaping flux.
The difference in flux between the two models provide a direct measure of how sensitive the predicted signal is to geometry. Large differences indicate optically thick conditions, while convergence indicates transparency.

\subsection{Thermal Blackbody Background}
The first detectability threshold is the thermal radiation field of the ejecta. We model this as a Compton-scattered thermal continuum at the local tracer temperature $T(t)$. 
The Planck photon number density per photon energy is:
\begin{equation}
n_{\rm Planck}(E, T) = \frac{8\pi}{(hc)^3} \cdot \frac{E^2}{e^{E / k_B T} -1},
\end{equation}
with units of $\rm photons \cdot \rm MeV^{-1} \cdot  \rm cm^{-3}$ when $E$ is in MeV. 
The Compton opacity per unit mass is computed from the energy-dependent Klein-Nishina total cross section \citep{Longair},
\begin{equation}
\sigma_{\rm KN}=\frac{3\sigma_T}{4}\left[\frac{1+x}{x^3}\left(\frac{2x(1+x)}{1+2x}-\ln(1+2x)\right)+\frac{\ln(1+2x)}{2x}-\frac{1+3x}{\left(1+2x\right)^2}\right],
\end{equation}
where $\sigma_T =  6.652 \times 10^{-25} \rm cm^2$ is the Thompson cross section, and $x = E/(m_e c^2)$ with $m_e c^2 = 0.511$ MeV. 
Then, the Compton opacity is:
\begin{equation}
\kappa(E) = \sigma_{\rm KN}(E) \cdot (Z/A)\cdot \frac{1}{m_u},
\end{equation}
in units of $\rm cm^2 \cdot g^{-1}$.
The factor $1/m_u$ (with $m_u = 1.67 \times 10^{-24}$g the nucleon mass), converts from per-baryon abundance to per-gram emission rate.
For r-process ejecta we adopt the electron-to-baryon ratio $Z/A=0.40$.

The corresponding thermal emission rate per unit mass is
\begin{equation}
B_{\rm bb}(E, T) = \kappa (E) \cdot c \cdot n_{\rm Planck}(E, T),
\end{equation}
with units of $\rm photons  \cdot \rm MeV^{-1} \cdot \rm g^{-1} \cdot \rm s^{-1}$.
Note that this quantity depends only on $T$ and $E$ because the local density cancels. 
The thermal background dominates only at very early times because as the ejecta cools, the tail of the Planck spectrum falls exponentially as $\exp(-E/k_BT)$.
Therefore, while the thermal background sets the earliest possible emergence time of the isomeric lines, it is negligible during the hours-to-week window relevant for $\gamma$-ray spectroscopy.

\subsection{Isomeric Gamma-ray Signal}
We now compute the $\gamma$-ray spectrum produced by isomeric de-excitation, following \citep{arxiv:2509.00267, arxiv:2510.08560}. The isomer-resolved PRISM network gives the abundance $Y_m(t)$ of each metastable species at each timestep and for each trajectory. Since PRISM abundances are per baryon, the transition flow rate per unit ejecta mass is
\begin{equation}
F_m(t) = Y_m(t) \cdot \Lambda_{mg} \cdot \frac{1}{m_u},
\end{equation}
in units of $\rm g^{-1} \cdot \rm s^{-1}$, where $\Lambda_{mg}$ is the temperature-dependent effective transition rate from the metastable to the ground state, computed as described in \S\ref{sec:theory}.

Each isomeric transition produces one photon at the de-excitation energy $E_{\gamma}^m$.
We represent this line on a discrete energy grid of width $\Delta E$
\begin{equation}
s_m(E) = \left\{
\begin{array}{l}
\frac{1}{\Delta E} \quad \texttt{if} \quad E \in \left ( E^m_{\gamma}  - \frac{\Delta E}{2}, E^m_{\gamma} + \frac{\Delta E}{2} \right ), \\
0 \quad \texttt{otherwise}.
\end{array}
\right.
\end{equation}
The resulting per-gram emission spectrum for one isomer is
\begin{equation}
S_{\gamma}^m(E, t) = F_m(t) \cdot s_m(E),
\end{equation}
with units of $\rm photons \cdot \rm MeV^{-1} \cdot \rm g^{-1} \cdot \rm s^{-1} $.
We adopt $\Delta E = 1$ keV, matched to the spectral resolution of COSI and well below that of AMEGO and LOX. 

For each isomeric transition at energy $E_\gamma$, we construct the summed observable flux by combining contributions from all tracers:
\begin{equation}
F^{\rm m}_{\rm line}(E_{\gamma},t)=\frac {E_{\gamma} \Delta E}{4\pi D^2} \sum_i M_i  f_{\rm esc,i}(E_{\gamma}, t) S^{\rm m}_{\gamma,i} (E_{\gamma},t),
\end{equation}
where $D$ is the distance from the source to the observer, $M_i$ is the tracer mass, and $f_{\rm esc, i}$ is the escape fraction per tracer velocity $v_i$.
Thus, by multiplying the differential photon spectrum by the line energy and integrating over the corresponding energy bin, the $\gamma$-ray flux correspond to energy fluxes in units of $\rm MeV\cdot \rm cm^{-2}  \cdot \rm s^{-1}$.

The corresponding thermal background is evaluated at the same energy:
\begin{equation}
F_{\rm bb}(E_{\gamma},t)=\frac {E_{\gamma} \Delta E}{4\pi D^2} \sum_i M_i  f_{\rm esc,i}(E_{\gamma}, t) B_{\rm bb,i} (E_{\gamma},t).
\end{equation}

We then define the thermal emergence time as the earliest time at which the ratio $F^{\rm iso}_{\rm line}/F_{\rm bb} \ge 1$ is satisfied for a small number of consecutive timesteps, ensuring robustness against numerical fluctuations.
For an individual trajectory, the thermal-emergence ratio reduces to $S^{\rm iso}_{\gamma,i}/B_{\rm bb,i}$. For the observable region-integrated criterion, however, we use the corresponding mass- and escape-weighted ratio of summed fluxes over all trajectories.
This time marks the transition from a thermal spectrum to one where discrete $\gamma$-ray lines can be observationally resolved.
We find that the thermal emergence time remains relatively stable across all isomers and escape geometries, occurring between $\sim 0.3-0.7$ s.

\subsection{Beta-decay Continuum and Observability}
After the thermal blackbody photon energy drops below the line emission, the relevant background is the $\gamma$-ray continuum from $\beta$-decay of other r-process nuclei. 
This continuum includes $\gamma$ cascades from excited daughter states, internal conversion followed by bremsstrahlung, and the overlap of many individual decay lines.

We compute this background directly from the PRISM output at each timestep, following the same algorithm as in  \citet{arxiv:2509.00267, arxiv:2510.08560}. 
This calculation uses the same trajectories, composition, and temperature-dependent effective-rate method as the isomer calculation, but sums over the $\beta$-decay $\gamma$ channels instead of the isomeric de-excitation channels.
The $\beta$-decay continuum spectral densities $S_{\gamma,i}^{\beta}(E, t)$ are read directly in units of $\rm photons \cdot MeV^{-1} \cdot g^{-1} \cdot s^{-1}$. 
Although the $\beta$-decay continuum is tabulated on a logarithmic energy grid, it is stored as a spectral density. We therefore interpolate the continuum to the isomeric line energy $E_{\gamma}$ and integrate it over the same comparison bandwidth used for the line flux.
The resulting local continuum flux is:
\begin{equation}
F_{\beta}(E_{\gamma},t)=\frac {E_{\gamma} \Delta E}{4\pi D^2} \sum_i M_i  f_{\rm esc,i}(E_{\gamma}, t) S_{\gamma,i}^{\beta}(E_{\gamma}, t).
\end{equation}
This gives the local $\beta$-decay continuum level at the exact line energy, against which the isomeric line is compared. 

For each trajectory and timestep, we compute: the isomeric line spectrum $S_{\gamma}^{\rm m}$, the $\beta$-decay continuum $S_{\gamma}^{\beta}$, and the thermal background $B_{\rm bb}$. 
A line may be observable only when it exceeds both the thermal background and the $\beta$-decay continuum and is above the detector sensitivity thresholds.
We use a fiducial Galactic event at a distance D=15 kpc and compare the predicted energy fluxes with representative detector sensitivity thresholds for current and proposed MeV $\gamma$-ray instruments (COSI, AMEGO, LOX). 
For each isomeric line, a detector is considered as sensitive only if the line energy lies within the detector’s energy range and the peak post-emergence energy flux exceeds the quoted threshold \citep{arxiv:1908.04334, arxiv:2101.03105}.
Specifically, we adopt energy-flux sensitivities of $2\times 10^{-5}$, $5\times 10^{-7}$ and $10^{-6}$ $\rm MeV \rm s^{-1} \rm cm^{-2}$ for COSI, AMEGO, and LOX, respectively.
These comparisons use nominal line sensitivity thresholds and do not model detailed detector response, exposure-time dependence, background rejection, or line broadening.

\subsection{Detectability of Isomer $\gamma$-Rays}

\begin{figure}[ht!]
\plotone{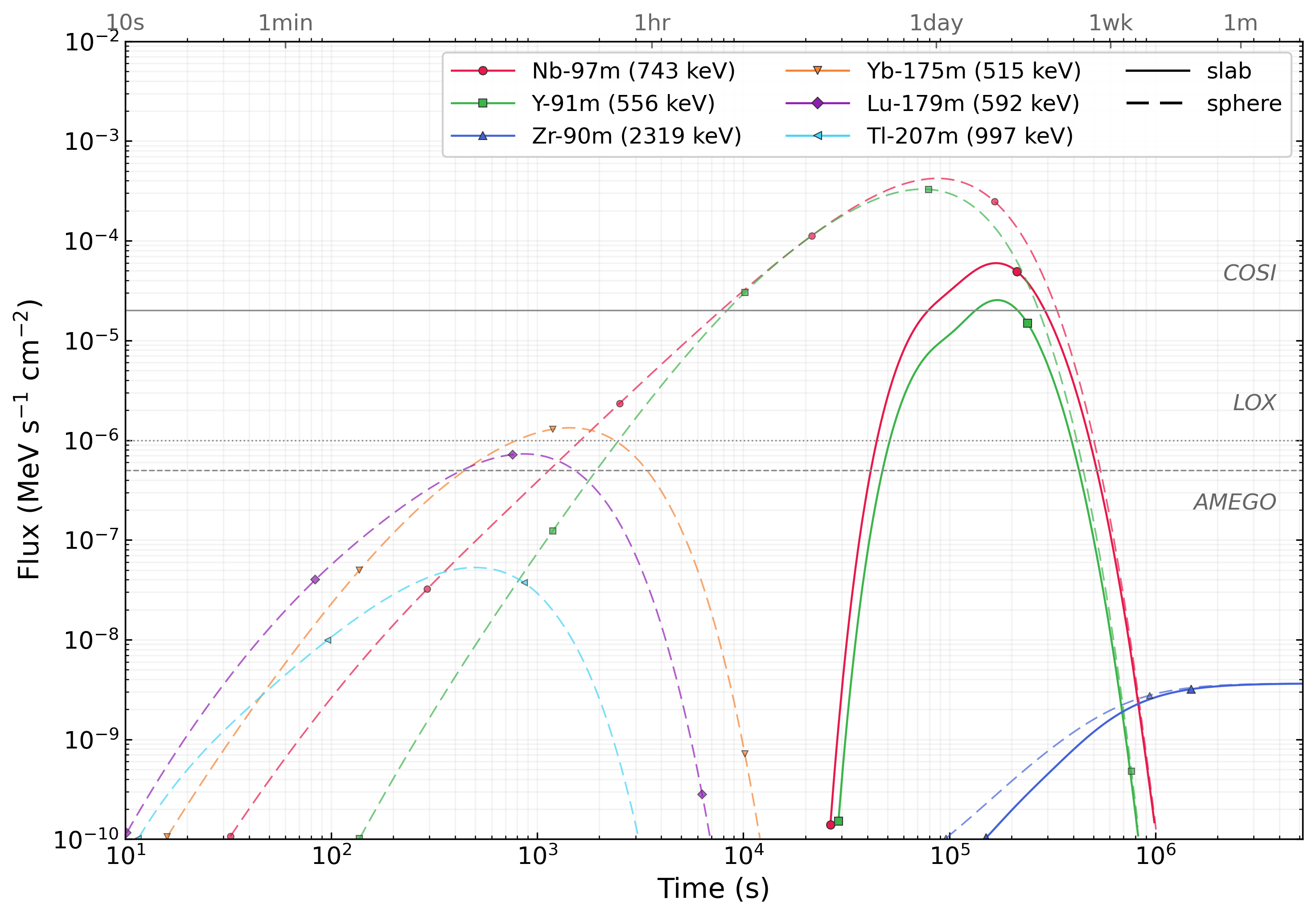}
\caption{Observable $\gamma$-ray energy fluxes for the individual isomers at $D=15$ kpc. Solid curves correspond to the slab and dashed to the sphere escape prescriptions. Horizontal lines indicate detector sensitivities and vertical orange lines the thermal emergence time.\label{fig:Figure2}}
\end{figure}

Figure \ref{fig:Figure2} shows the incident flux from isomeric transitions at a fiducial distance of $D=15$ kpc for both escape geometries as compared to detectability estimates, and Table \ref{tab:detection} summarizes the corresponding post-emergence peak fluxes and detectability estimates. We find the flux of Nb-97m and Y-91m to be above detectability thresholds in both geometries for all detectors, and Yb-175m and Lu-179m to be slightly above detectability thresholds for AMEGO in the spherical geometry, but not detectable in the slab geometry. Peak $\gamma$-ray emission occurs during the transition from optically thick to optically thin conditions where the escape fraction is strongly geometry-dependent, and consequently, different escape prescriptions produce order-of-magnitude variations in the peak flux and the time which the peak occurs. This emphasizes the importance of careful modeling of the ejecta and escape of $\gamma$-rays.

\begin{figure}[ht!]
\plotone{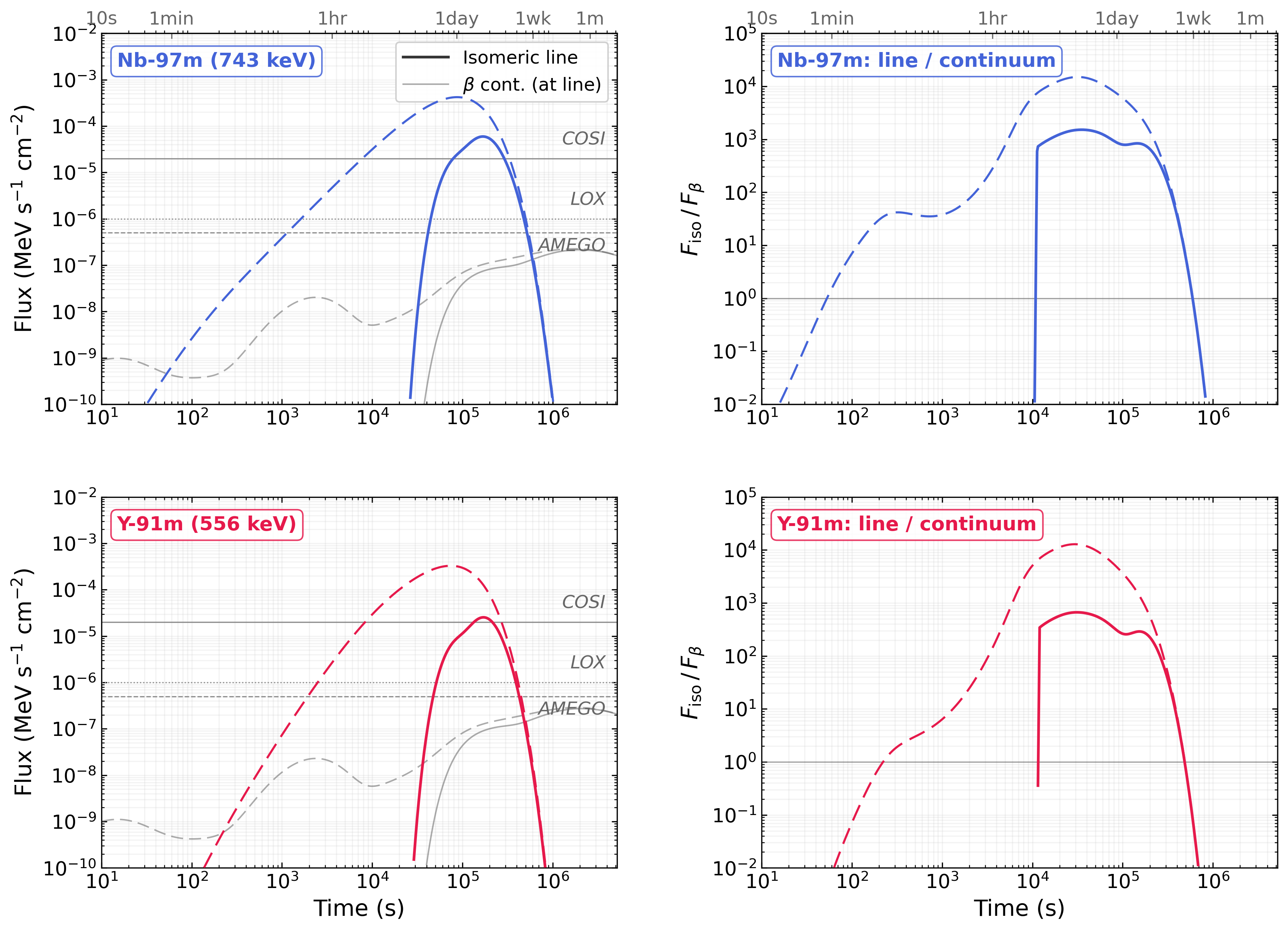}
\caption{Observable $\gamma$-ray energy fluxes for Nb-97m and Y-91m at $D=15$ kpc. The left panels compare the delayed isomeric line flux with the local $\beta$-decay continuum at the same energy and with detector sensitivities. The right panels show the ratio between the line and continuum fluxes. Solid curves correspond to the slab and dashed to the sphere escape prescriptions.
Horizontal lines indicate detector sensitivities and vertical orange lines the thermal emergence time.\label{fig:Figure3}}
\end{figure}

Of the isomers we have considered, Nb-97m and Y-91m are the most promising candidates for observation. In Figure \ref{fig:Figure3}, we compare the isomer flux of these nuclei to the $\gamma$-ray emission from other $r$-process nuclei. We find that both the 743 keV line from Nb-97m and 556 keV line from Y-91m are the dominant contributors at their corresponding energy, and therefore may be directly observable in $\gamma$-ray detectors. However, we emphasize that these calculations are intended to demonstrate the potential importance of isomeric contributions to the $\gamma$-ray signal, not to provide a rigorous calculation of the observability. Such a calculation would require consideration of the entire ejecta, not just a representative sample of trajectories, and would also require detailed radiation transport calculations.

\begin{deluxetable*}{rlllll}[ht!]
\tablecaption{Region-integrated peak fluxes and times of candidate isomeric gamma-ray lines at 15 kpc}
\label{tab:detection}
\tablehead{
\colhead{Isomer} & \colhead{Geometry} & \colhead{$t_{\rm emerge}$ [ms]} & \colhead{Peak Flux $[\rm MeV \rm s^{-1} \rm cm^{-2}$]} & \colhead{Peak Time} & \colhead{Detectability}
}
\startdata
Nb-97m & Sphere & $572.6$ & $4.22{\times}10^{-4}$ & $1.0$ day & COSI, AMEGO, LOX \\
{    } & Slab   & $585.4$ & $5.97{\times}10^{-5}$ & $2.0$ days & COSI, AMEGO, LOX \\
\hline
Y-91m & Sphere & $704.1$ & $3.29{\times}10^{-4}$ & $20.4$ hr. & COSI, AMEGO, LOX  \\
{   } & Slab   & $713.6$ & $2.54{\times}10^{-5}$ & $2.0$ days & COSI, AMEGO, LOX \\
\hline
Zr-90m & Sphere & $331.6$ & $1.72{\times}10^{-9}$ & $2$ months &  -- \\
{   }  & Slab   & $343.0$ & $3.62{\times}10^{-9}$ & $2$ months &  -- \\
\hline
Yb-175m & Sphere & $694.5$ & $1.33{\times}10^{-6}$ & $23.8$ min & AMEGO, LOX \\
{    }  & Slab   & $713.6$ & $< 10^{-10}$ & -- & -- \\
\hline
Lu-179m & Sphere & $627.6$ & $7.29\times 10^{-7}$ & $14.4$ min & AMEGO \\
{    }  & Slab   & $652.6$ & $< 10^{-10}$ & -- & -- \\
\hline
Tl-207m & Sphere & $496.6$ &  $5.29{\times}10^{-8}$ & $8.2$ min & -- \\
{    }  & Slab   & $517.8$ & $< 10^{-10}$ & -- & -- \\
\enddata
\end{deluxetable*}

\section{\label{sec:discussion}Discussion}

We have found that thermal transitions from nuclear isomers can produce significant $\gamma$-ray lines which may be observable in current and future $\gamma$-ray observatories. This represents a challenge to the conventional methodology of $\gamma$-ray emission modeling. Typical calculations convolve a ground-state only reaction network with individual decay spectra, usually as given in ENDF or ENSDF. These decay spectra are difficult to correctly couple to ground-state reaction networks, as the decay of the nuclear isomer is tabulated separately. For example, even though Nb-97m is populated $>99\%$ of the time in the $\beta$-decay of its parent Zr-97, the 743 keV line is not present in the Zr-97 decay spectra, and is therefore missing from current ground-only $\gamma$-ray emission modeling. Y-91m and Nb-97m serve as representative examples of potential isomeric contributions. We expect that consideration of all known isomers will reveal additional examples.

The dynamical treatment of nuclear isomers may have additional effects on the $\gamma$-ray spectra that have not been realized in this work. First, the timescale of $\gamma$-ray emission from isomeric transitions is determined by the isomer lifetime rather than the lifetime of the parent nucleus, which is especially important when the isomer is much longer lived than its parent. This is not the case for the isomers we found to be significant: Y-91m ($T_{1/2} = 49.7$ min, $T_{1/2}^{\rm parent} = 9.65$ hr) or Nb-97m ($T_{1/2} = 58.7$ s, $T_{1/2}^{\rm parent} = 16.7$ hr). However, for other nuclei, such as Te-127m, ($T_{1/2} = 106.1$ days, $T_{1/2}^{\rm parent} = 3.85$ days), such effects may be significant if the isomer is populated during $\beta$-decay. In addition, such long-lived isomers may affect the decay timescale of the corresponding ground state ($T_{1/2}^{\rm Te-127} = 9.35$ hr), which could also have observational effects.

Second, the $\beta$-feeding into different states of the daughter nucleus can change as a function of temperature, thereby changing the ratio at which the ground and isomeric states populate. We have explored the effect of explicit temperature dependence on effective transition rates (see Appendix \ref{sec:sensitivity}) and found a significant temperature dependence. This temperature dependence is unlikely to have a significant effect on timescales for direct $\gamma$-ray observation, as the ejecta has cooled significantly by later times when the isomers we have identified (Y-91m , Nb-97m) are produced. However, it may have significant effects on isomers produced at earlier timescales, which could have larger implications for kilonova physics. Finally, isomers may directly $\beta$-decay, producing an entirely different $\gamma$-ray spectra. 

These results demonstrate that systematic, dynamical treatment of nuclear isomers is essential to the accurate modeling of the $\gamma$-ray emission spectra. Such a calculation presents its own challenges, particularly for nuclei where nuclear data is incomplete or uncertain. We have found that the predicted line fluxes depend sensitively on (i) $\beta$-feeding branching fractions to ground and isomeric states and (ii) direct transition rates between all states. These quantities are impacted by the details of the level structure, including excitation energies, spins, parities, and lifetimes. 

\section{\label{sec:conclusion}Conclusion}

In this work, we investigated nuclear isomers as sources of observable $\gamma$-ray line emission in the r-process ejecta of binary neutron star mergers. Using an isomer-resolved extension of the PRISM nucleosynthesis network applied to 30 post-merger trajectories, we evolved the time-dependent populations and observable $\gamma$-ray fluxes of metastable nuclei under temperature-dependent $\beta$-feeding conditions. From an initial set of 12 candidates, we have identified two isomeric transition lines: the 743.3 keV line of Nb-97m and 555.6 keV line of Y-91m, which dominate the observable emission hour-to-day timescales at their respective energies, producing fluxes within the detectability range of proposed MeV $\gamma$-ray missions for Galactic events. 

These results demonstrate that isomeric $\gamma$-ray emission may produce a distinct observational signal for r-process nucleosynthesis. Future $\gamma$-ray observatories are projected to detect $\gamma$-rays from $r$-process events, and the accurate characterization of these observations will require reaction networks that evolve isomeric states dynamically.

\begin{acknowledgments}
We thank Kelsey Lund for providing the nucleosynthesis tracers, as well as insightful discussions on their appropriate usage.
This research was made possible in part by the NSF REU Site: Appalachian Mathematics and Physics Award No. 2349289 to Marshall University. 
MCBH also acknowledges support from the NSF grant No. PHY-1748958 to the Kavli Institute for Theoretical Physics.
Research presented in this article was supported by the Laboratory Directed Research and Development program of Los Alamos National Laboratory under project number 20240004DR.  
LANL is operated by Triad National Security, LLC, for the National Nuclear Security Administration of U.S. Department of Energy (Contract No. 89233218CNA000001). 
\end{acknowledgments}

\appendix

\section{\label{sec:validation}Validation Against Al-26m}
\begin{figure}[ht!]
\plotone{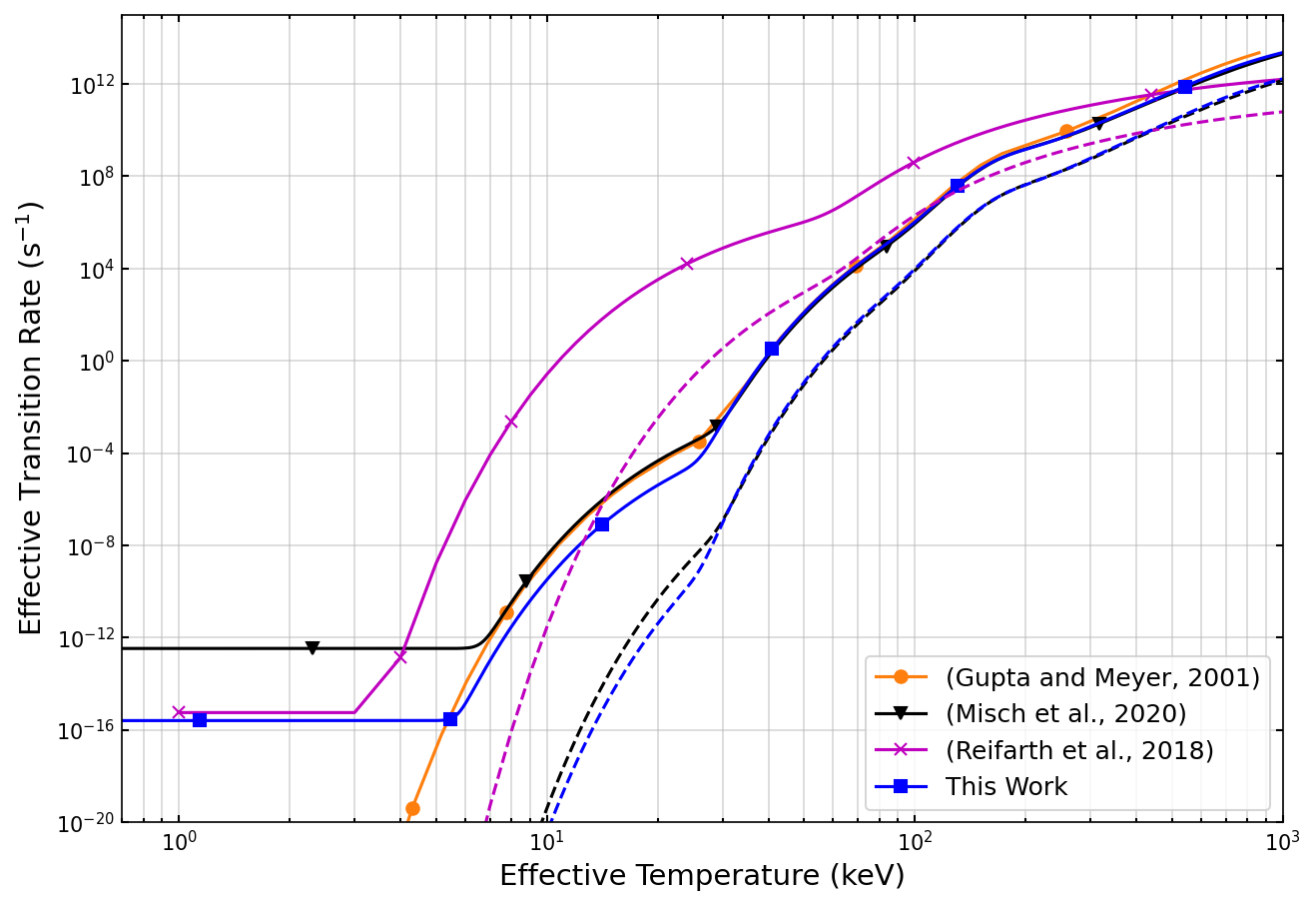}
\caption{A comparison of effective transition rates for Al-26m between this work (blue squares), \citep{arxiv:2010.15238} (black triangles), \citep{meyer} (orange circles), and \citep{arxiv:1803.08335} (magenta x's). Solid lines represent the isomer-to-ground transition rates, while dashed lines represent ground-to-isomer transition rates.
\label{fig:Figure4}}
\end{figure}

As a first application, we selected the isomer Al-26m, which is well characterized in the literature due to its astrophysical relevance \citep{meyer, arxiv:2010.15238, arxiv:1803.08335}.
The calculated rates, both from ground to isomer and from isomer to ground, are shown in Figure \ref{fig:Figure4} as a function of effective temperature (units of keV), along with previous rates from literature. We find we are in good agreement with the work of \citep{meyer} and \citep{arxiv:2010.15238}, especially at high temperatures.

\begin{figure}[ht!]
\plotone{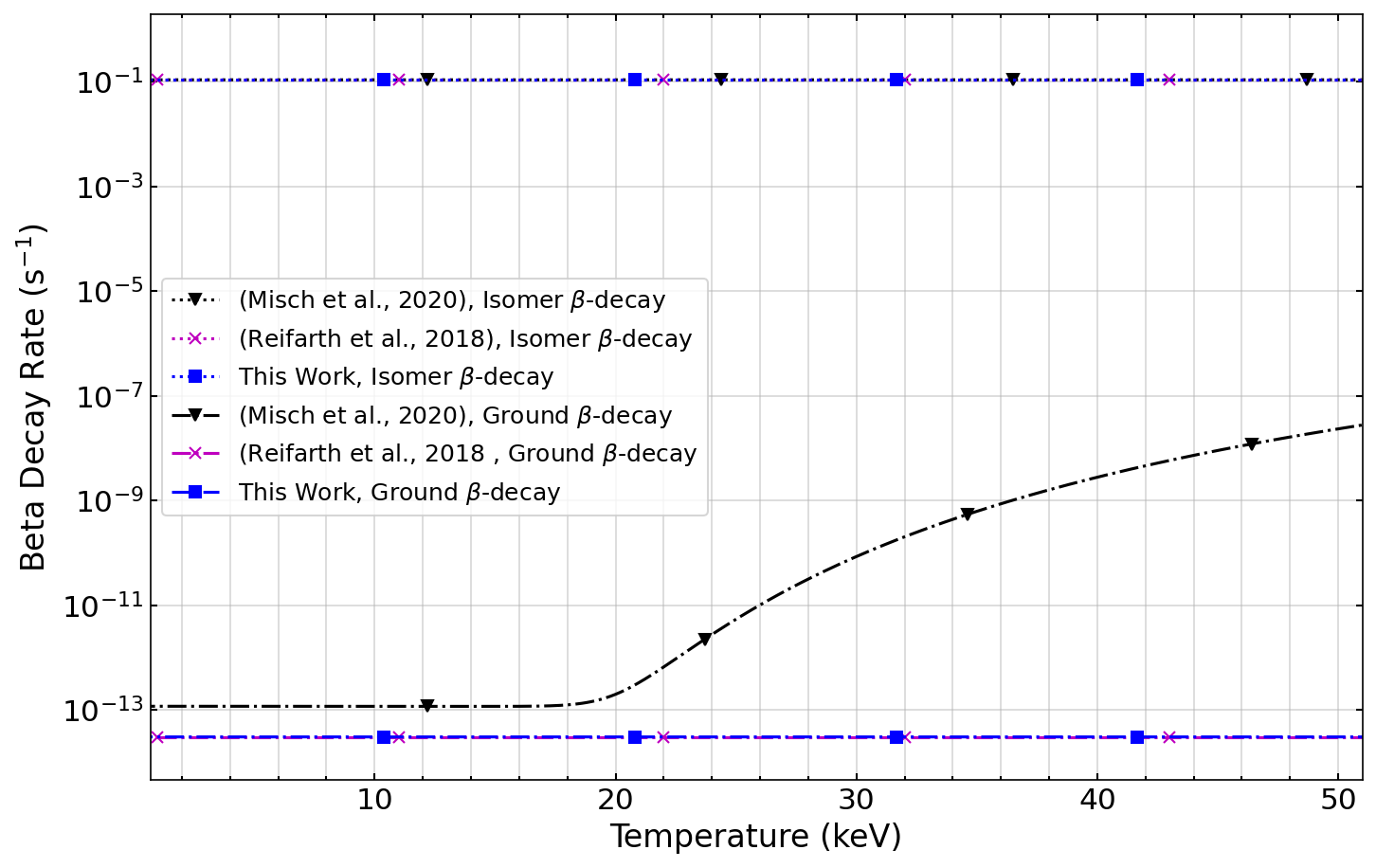}
\caption{A comparison of $\beta$-decay rates for Al-26m between this work (blue squares), \citep{arxiv:2010.15238} (black triangles), and \citep{arxiv:1803.08335} (magenta x's). Dotted lines denote decay from the isomeric state, and dash-dot lines denote decay from the ground state.
\label{fig:Figure5}}
\end{figure}

We also validate the implementation of (eq.\ref{eq:beta}) to find the temperature-dependent $\beta$-decay rates of both the ground and isomer states of Al-26m (Figure \ref{fig:Figure5}). 
Our $\beta$-decay rates from both the ground and isomeric states are in good agreement with \citep{arxiv:1803.08335}. For the ground-state $\beta$-decay rate, we find some tension with \citep{arxiv:2010.15238} at high temperatures.

We initially tested an experiment-only implementation using measured $\gamma$-transition data, which failed to reproduce the literature benchmark.  
Including Weisskopf estimates for the unmeasured transitions between intermediate levels recovered agreement with the published result. This demonstrates that experimental data alone is insufficient even for the well-studied Al-26m case and motivates the inclusion of theoretical estimates throughout our calculations.

\section{\label{sec:application}Application to Selected Isomers}

Having validated the formalism against Al-26m, where measured and computed rates agree across the relevant temperature range, we now apply the formalism to the relevant set of isomers introduced in Section \ref{sec:selection}.
We compute (i) the effective m$\leftrightarrow$g transition rates as a function of temperature, and 
(ii) the ratio of $\beta$-decay feeding into the isomeric versus ground state. 
The first quantity provides the lifetime of the isomer in the ejecta, while the second influence its population once $\beta$-feeding from longer-lived parents becomes the dominant source.

\begin{figure}[ht!]
\includegraphics[width=0.75\linewidth]{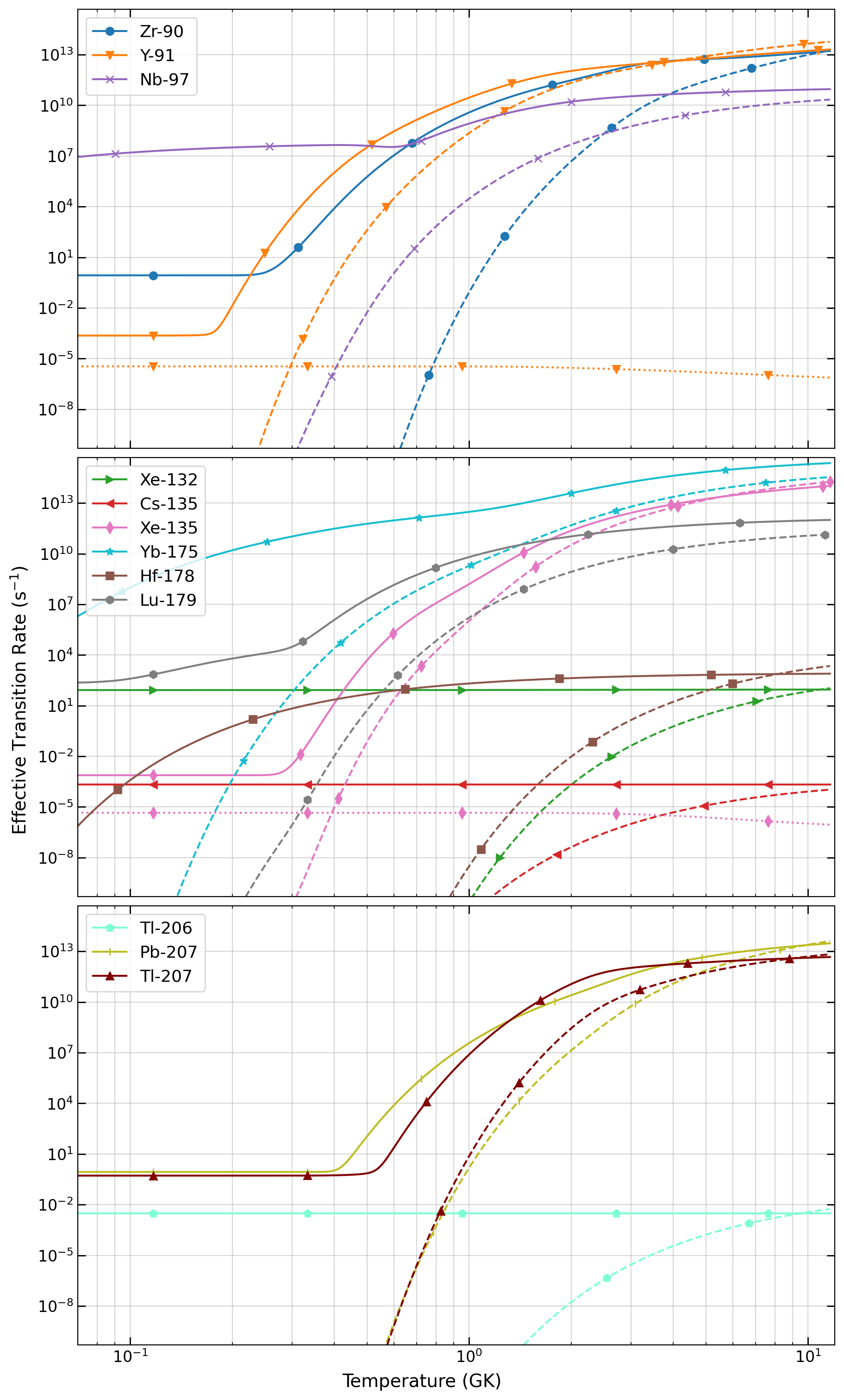}
\caption{Effective transition rates as function of temperature for selected isomers. Solid lines denote isomer-to-ground rates, dashed lines ground-to-isomer rates, and dotted lines isomeric $\beta$-decay rates. Top: isomers around $1^{st}$ r-process peak. Middle: isomers around $2^{nd}$ r-process and rare-Earth peaks. Bottom: isomers around $3^{rd}$ r-process peak.
\label{fig:Figure6}}
\end{figure}

Figure \ref{fig:Figure6} shows the effective transition and $\beta$ decay rates as a function of temperature for the selected isomers,  grouped by r-process peak. Three features are robust across the sample:
At low temperatures, the m$\rightarrow$g rate dominates over the g$\rightarrow$m rate by many orders of magnitude. In this regime the isomer behaves as a metastable state with a fixed spontaneous lifetime, and the ground state acts as a one-way sink.
Rising temperature thermally populates intermediate levels that bridge the m and g states, driving both rates upward rapidly. For most isomers in the sample, the rates rise by several orders of magnitude.
At sufficiently high temperatures, depending on the isomer, the rates converge, signaling thermal equilibrium between the two states. The temperature at which equilibrium sets in depends on the energy gap between m and g and on the presence of bridging levels at intermediate energies. 
We note that direct $\beta$-decay from the isomer is a sub-dominant channel for nearly all isomers in our sample.

\begin{figure}[ht!]
\plotone{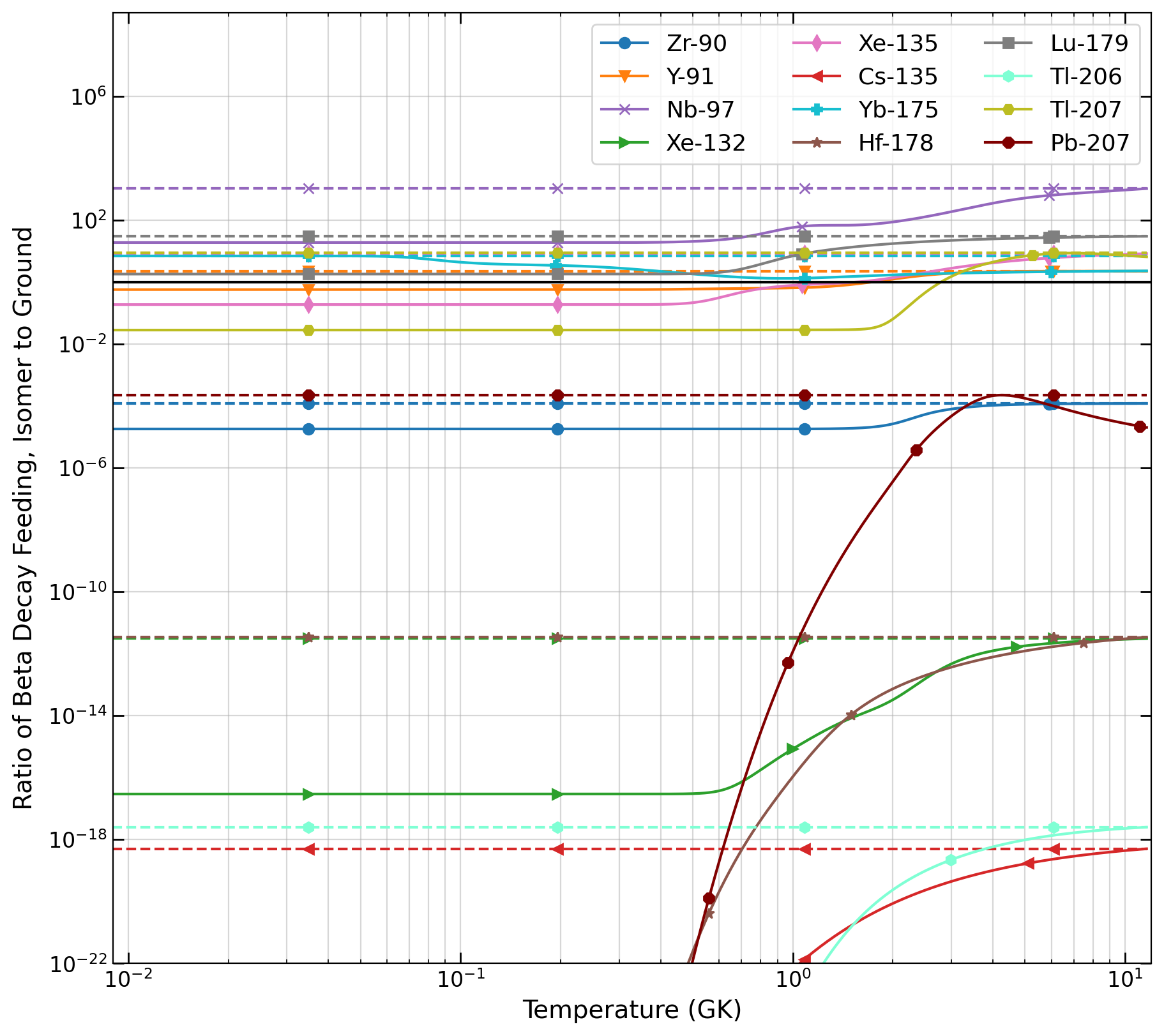}
\caption{Ratio of $\beta$-decay feeding rates, isomer to ground, for twelve different isomers, as a function of temperature. Solid curves show the full temperature dependent behavior, while the dashed lines represent the maximum (optimal) ratio value. The black solid line denotes a ratio of 1, with equal transitions to ground and isomeric states from parent $\beta$-decay.
\label{fig:Figure7}}
\end{figure}

Figure \ref{fig:Figure7} shows the ratio of $\beta$-feeding into the isomeric versus ground state ($P_m /  P_g$), as a function of temperature for the same set of isomers.  
To our knowledge, this temperature-dependent treatment of $\beta$-feeding to isomeric states in an r-process context has not been computed before and no direct literature comparison exists for these quantities.
We validated the implementation through probability conservation at all temperatures, implementing a conservative treatment of disconnected levels. 
Namely, for $\beta$-populated levels lacking complete $\gamma$-cascade data connecting them to the ground or isomeric state, we assign 100\% feeding probability to the ground state instead of excluding that level. 

The implication for the detectable signal is that the isomer population at any given time is determined by a balance between three temperature-dependent quantities: the m$\leftrightarrow$g rate, the $\beta$-decay rate (Figure \ref{fig:Figure6}), and the $\beta$-feeding ratio to the isomer (Figure \ref{fig:Figure7}). Treating any one of these as temperature-independent biases the result by orders of magnitude.

\section{\label{sec:sensitivity}Sensitivity to the Treatment of Beta-feeding}
\begin{figure}[ht!]
\plotone{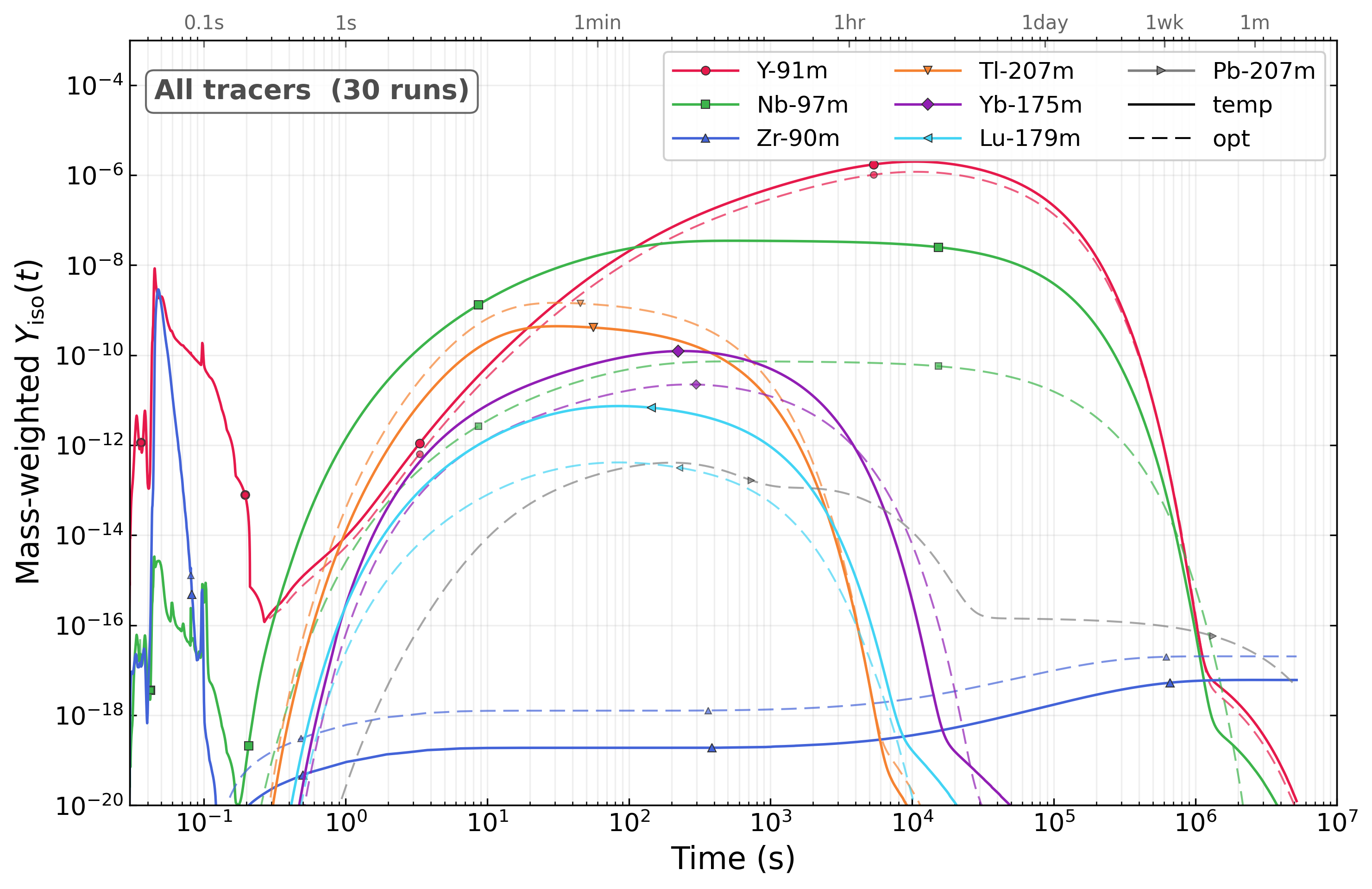}
\caption{Isomeric abundances for temperature-dependent and optimal $\beta$-feeding treatments. Sustained $\beta$-fed isomers show treatment-dependent enhancements (Nb-97m) or suppression (Pb-207), while the prompt component remains unchanged.
\label{fig:Figure8}}
\end{figure}

To assess the sensitivity of the predicted isomeric abundance to the treatment of parent-to-daughter $\beta$-feeding, we compare a temperature-dependent  implemented through \texttt{prob-rxn} entries in PRISM against an alternative calculation in which the $\beta$-feeding fractions are frozen at their high-temperature values and implemented as \texttt{prob-decay} entries. 
The two treatments produce four distinct behaviors across the viable isomer set, which are visible in Figure \ref{fig:Figure8}. We regard the temperature-independent treatment as physically inadequate for the conditions relevant to neutron-star merger ejecta, and include the comparison primarily for methodological transparency rather than as an alternative physical prediction.  

For Y-91m, Nb-97m, Yb-175m, and Lu-179m, the temperature-dependent treatment enhances the sustained isomeric population relative to constant feeding. 
The magnitude of the enhancement depends sensitively on the daughter-level structure and the redistribution of thermal population among excited states.

Tl-207m exhibits the opposite behavior. In this case, the temperature-dependent treatment suppresses the sustained isomeric population by a factor of $\sim 3.4$ relative to constant feeding ($2.7\times10^{-9}$ vs. $9.2\times 10^{-9}$).
The isomer remains above the production threshold and therefore stays within the viable set, but the predicted line flux is correspondingly reduced. 
In both the enhanced and suppressed cases, the underlying mechanism is the same: thermal redistribution modifies the effective $\beta$-feeding among daughter levels, and subsequent cascades can either increase or decrease the net feeding into the metastable state.

The most consequential treatment-dependent effect is the elimination of Pb-207m from the viable set. Under constant feeding, the Tl-208 $\rightarrow$ Pb-207m channel produces populations above the adopted production threshold in several trajectories. 
Under the temperature-dependent treatment, however, thermal redistribution shifts the effective feeding preferentially toward the ground state of Pb-207, suppressing the metastable population below threshold in all trajectories. The choice of $\beta$-feeding prescription therefore changes not only the predicted magnitudes of the observable fluxes, but also the identity of the viable observable-isomer set itself.

Finally, a qualitatively different behavior is seen in the prompt population during active r-process burning (Zr-90m, Y-91m, and Nb-97m), which is identical under both feeding treatments. These isomers are populated directly by r-process flow through neutron capture and photodissociation channels into the metastable state, rather than through delayed $\beta$-feeding. 
This prompt component is physically distinct from the later sustained $\beta$-fed accumulation.
The agreement of the flash-dominated isomers between the two treatments at very early times, combined with the substantial late-time differences for the same nuclei, shows that the temperature-dependent treatment primarily affects the sustained $\beta$-fed populations relevant during the kilonova observation window. Its influence is smaller at the earliest times, when the direct dynamical r-process flow dominates the isomeric population.

\bibliography{AASisomers}{}
\bibliographystyle{aasjournalv7}

\end{document}